\documentclass[
]{ceurart}

\usepackage{listings}
\usepackage[nolist]{acronym}
\begin{document}

\begin{acronym}[]
    \acro{AI}{Artificial Intelligence}
    \acro{AUI}{Attentive User Interface}
    \acro{HMI}{Human-Machine Interface}
    \acro{IVIS}{In-Vehicle Information System}
    \acro{LLM}{Large Language Model}
    \acro{SA}{Situation Awareness}
    \acro{TOR}{Take-Over Request}
    \acro{V2X}{Vehicle-to-Everything}

\end{acronym}

\copyrightyear{2024}
\copyrightclause{Copyright for this paper by its authors.
  Use permitted under Creative Commons License Attribution 4.0
  International (CC BY 4.0).}

\conference{MUM'23 Workshop on Interruptions and Attention Management: Exploring the Potential of Generative AI, December 3, 2023, Vienna, Austria}

\title{Generative AI and Attentive User Interfaces: Five Strategies to Enhance Take-Over Quality in Automated Driving}

\author[1]{Patrick Ebel}[%
orcid=0000-0002-4437-2821,
email=ebel@uni-leipzig.de,
url=https://ciao-group.github.io,
]
\address[1]{ScaDS.AI, Leipzig University,
  Humboldtstraße 25, 04105 Leipzig, Germany}

\begin{abstract}
As the automotive world moves toward higher levels of driving automation, Level 3 automated driving represents a critical juncture. In Level 3 driving, vehicles can drive alone under limited conditions, but drivers are expected to be ready to take over when the system requests. Assisting the driver to maintain an appropriate level of \ac{SA} in such contexts becomes a critical task. 
This position paper explores the potential of \acp{AUI} powered by generative \ac{AI} to address this need. Rather than relying on overt notifications, we argue that \acp{AUI} based on novel AI technologies such as large language models or diffusion models can be used to improve \ac{SA} in an unconscious and subtle way without negative effects on drivers overall workload. Accordingly, we propose 5 strategies how generative \acp{AI} can be used to improve the quality of takeovers and, ultimately, road safety.
\end{abstract}


\maketitle
\section{Introduction}
The advent of automated driving is changing the transportation landscape. The first cars with Level 3~\cite{.SAEJ3016Taxonomy.2021} driving automation features are on public roads~\cite{Mercedes-Benz.ConditionallyAutomated.2021} and many more will follow. While the purely technical components are becoming more sophisticated, critical issues regarding the interaction between humans and automation have yet to be resolved. \acp{TOR} emerge as a key component in this evolution. In Level 3 automated driving, the automated driving features can drive the vehicle under limited conditions, and drivers are relieved of the constant obligation to monitor the driving environment~\cite{.SAEJ3016Taxonomy.2021}. They can play with their mobile phones, interact with in-vehicle infotainment systems, or focus on conversations with their passengers. In other words, drivers can become disengaged from the driving task and the driving environment even though they must take over control once the car requests so. This presents a unique challenge: when a TOR is initiated, a disengaged driver is thrust back into a control role, often under conditions that require rapid comprehension and action.

Current research shows that engagement in non-driving activities, and thus loss of awareness of the driving environment, can reduce the quality of driver takeovers~\cite{McDonald.ComputationalSimulations.2019, Vogelpohl.TransitioningManual.2018}. Therefore, it is crucial to redirect the driver's attention to the road in a timely manner. While the question of how to assist drivers in maintaining or restoring sufficient SA has not been definitively answered~\cite{Marti.ImpactDriver.2022}, research suggests that sudden warnings aimed at redirecting the driver's attention often have the unintended side effect of increasing workload~\cite{Ma.TakeGradually.2021}. This increase in workload and mental stress can, in turn, lead to a decrease in take-over performance~\cite{Agrawal.EvaluatingImpacts.2021}. A seamless transition from automated to manual driving is therefore essential. 

But how can the transition from a state in which the driver can be fully disengaged from the driving task to a state in which the driver must be fully aware of the driving situation to handle a potentially dangerous driving task be made subtly and smoothly? \citet{DeGuzman.AttentiveUser.2022} point out that \acp{AUI}, that have been shown to effectively manage \ac{SA} in manual driving, can potentially also be beneficial for automated driving. Other recent work, for example by \citet{Wintersberger.AttentiveUser.2019}, underlines the potential of \acp{AUI} to improve take-over quality. In this position paper we go a step further and argue that in particular the combination of \acp{AUI} and generative AI technologies such as \acp{LLM} and Diffusion Models (e.g., Stable Diffusion~\cite{Rombach.HighResolutionImage.2022} or DALLE-3~\cite{Betker.ImprovingImage.2023}) can help to subtly bring the driver back into the loop or even subconsciously maintain the required level of \ac{SA}. When fine-tuned with the rich sensor data available in today's cars, these models can generate a comprehensive picture of the driving scenario and select guidance strategies tailored to the driving situation and the driver's state. Not only can they organically guide the driver back to control when the situation requires immediate control, they can also subtly enhance the driver's \ac{SA} in situations of increasing uncertainty, where it is not entirely clear whether a take-over will be issued. This prepares the driver without appearing overly cautious.

In the following we present five strategies that employ generative AI and in particular \acp{LLM} and Diffucion Models to serve as an inspiration for future research.

\section{Related Work}

In the following, we will give a brief overview of current research related to \acp{TOR} in general and the role that \acp{AUI} can play to improve \acp{TOR}.

\subsection{Take-Over Requests in Automated Driving}

In Level 3 automated driving, the automated driving functions can drive the vehicle under limited conditions~\cite{.SAEJ3016Taxonomy.2021}. In contrast to manual and assisted driving (L0-L2), the driver is relieved of the constant need to monitor the driving environment. However, the driver is required to be prepared to regain control in emergency situations, such as system failure or when the upcoming driving situation is outside the operational design domain of the system~\cite{Morales-Alvarez.AutomatedDriving.2020}. In these situations the automated driving systems triggers a \ac{TOR} notifying the driver to take over the driving task~\cite{.SAEJ3016Taxonomy.2021}. For such transfers of control back to the driver two scenarios need to be distinguished: \textit{``scheduled''} \acp{TOR} in situations in which the systems is aware of an upcoming \ac{TOR} (e.g., due to a highway exit or known road closure) and \textit{``imminent''} \acp{TOR} in sudden emergency situations (e.g., a broken down car blocking the road)~\cite{Wintersberger.AttentiveUser.2019}. While the latter is considered to be the most critical problem of Level 3 driving, it is unclear how often emergency \acp{TOR} are triggered~\cite{Wintersberger.AmDriving.2017, Eriksson.TakeoverTime.2017}, and it is assumed that as technology evolves (e.g., sensor range, \ac{V2X} communication), their frequency may decrease and the frequency of scheduled \acp{TOR} will increase. Accordingly, it is important that drivers are able to regain control and appropriate awareness of the driving situation such that they can handle the upcoming driving task safely. Related work shows that the reaction time to \acp{TOR} is an indicator for safety and \ac{TOR} quality~\cite{Wintersberger.AmDriving.2017, McCall.TaxonomyAutonomous.2019}. Studies on TOR quality further show that reaction time and driving performance are influenced by the driving context (e.g., road curvature~\cite{SadeghianBorojeni.ReadingDriving.2018} or traffic~\cite{Radlmayr.HowTraffic.2014}), driver behavior (e.g., engagement in secondary tasks~\cite{McDonald.ComputationalSimulations.2019, Gold.TakeHow.2013}, driver state (e.g., fatigue~\cite{Feldhutter.EffectFatigue.2019}), and \ac{TOR} modality (e.g., visual, vibrotactile, or auditory \cite{Yoon.EffectsTakeover.2019}).

These findings highlight that for safe takeovers, a holistic understanding of the current driving situation and the state of the driver is important to trigger context-dependent \acp{TOR}.

\subsection{Leveraging Attentive User Interface to Improve Take-Over Requests}

\acfp{AUI} are \textit{``computing interfaces that are sensitive to the user’s attention''}~\cite{Vertegaal.AttentiveUser.2003}. These interfaces therefore adapt the type and amount of information displayed based on the attentional state of the user and/or the attentional demands of the environment~\cite{DeGuzman.AttentiveUser.2022}. For example, due to the driver's current high stress level and the complex driving situation, an incoming call that's predicted to be of low urgency, may not be immediately put through, but rather suppressed until the driving situation allows it. Thus, \acp{AUI} can not only adjust the timing (e.g., as proposed by \citet{Wintersberger.LetMe.2018}) or the visual representation, but also consider the costs and benefits of conflicting actions by taking into account the driver's state and the driving situation~\cite{Braun.AffectiveAutomotive.2022}.

\citet{DeGuzman.AttentiveUser.2022} suggest that \acp{AUI}, that have been shown to effectively manage \ac{SA} in manual driving, may be also beneficial in automated driving. The authors identify several strategies for adapting UIs to either optimize attentional demand or to redirect the driver's attention to the road. However, they argue that only little research exists that studies the effect of \acp{AUI} in automated driving. One of the few studies that show the potential of \acp{AUI} for automated driving is presented by \citet{Wintersberger.AttentiveUser.2019} who argue that \acp{AUI} can improve take-over behavior. Their results show that \acp{AUI} improve driving performance, reduce the stress induced to drivers, and reduce the variance in the response times of scheduled \acp{TOR}.

\section{How Generative AI can Enhance TOR Quality}

\begin{figure}
  \includegraphics[width=\textwidth]{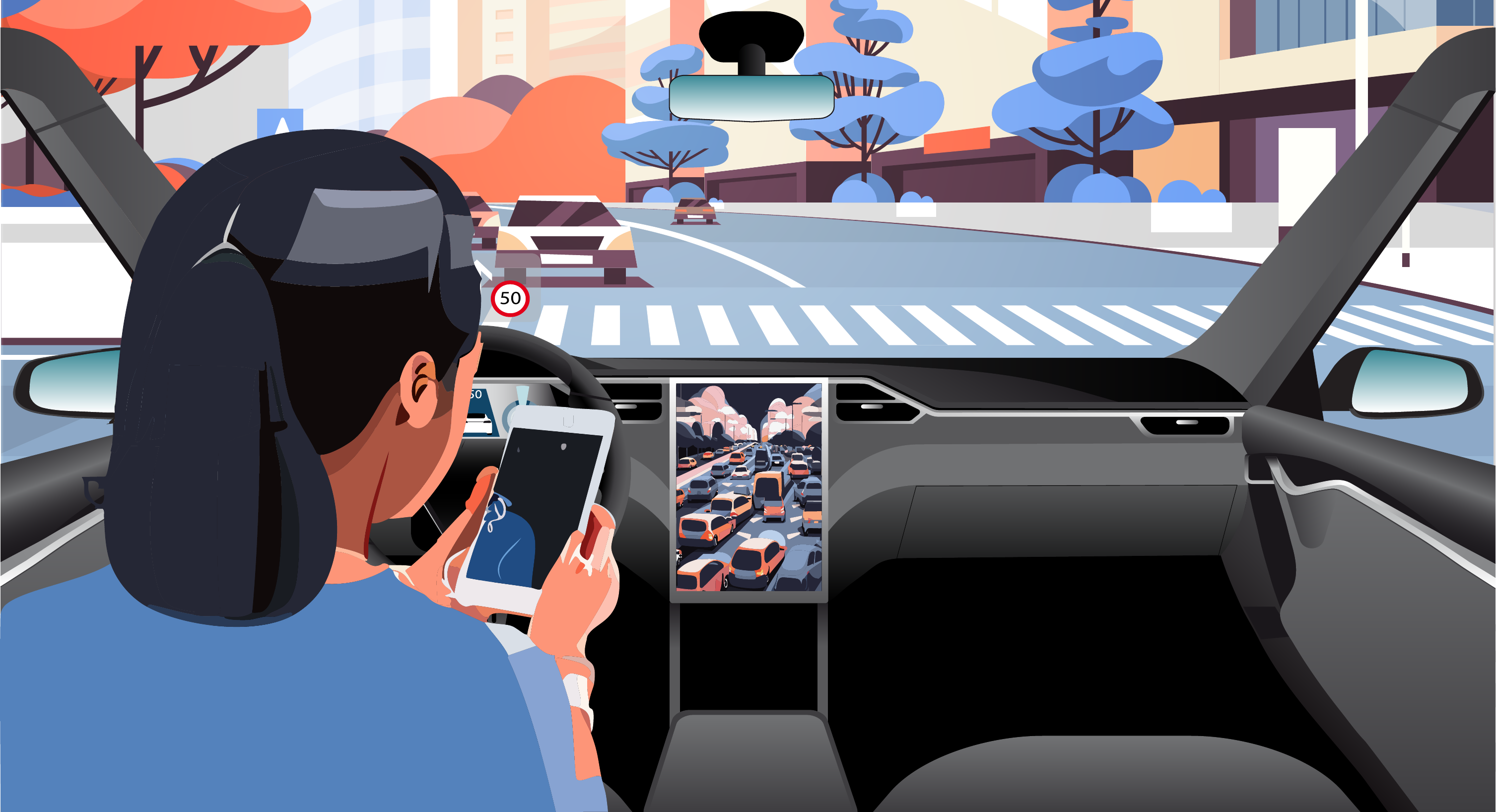}
  \caption{A hypothetical scenario: A person interacting with their mobile phone while driving in a Level 3 automated car. The current driving situation is under control and there is no reason to trigger a take-over request. However, the intelligent TOR assistant has detected a traffic jam ahead that may require the driver to take over. Knowing that the driver is engaged in a task on the smartphone, the TOR assistant decides to play an AI-generated video of the upcoming traffic situation on the center stack touchscreen. The driver will subconsciously recognize the moving scene on the center stack touchscreen and be more aware of the upcoming traffic scenario. The increased situation awareness will lead to a an increase in take-over quality.}
  \label{fig:teaser}
\end{figure}

To effectively tailor the interventions to the driving situation and the driver's state, an intelligent TOR agent needs access to the driving automation features, the car sensors (e.g., cameras and radar sensors, the cabin cameras) and access to the in-vehicle \acp{HMI} (e.g., infotainment system or head-up display). This information is already available in some modern production cars as shown in the works by~\citet{Ebel.Forces.2023, Ebel.ICEBOAT.2023}. To personalize interventions, it is also necessary to access personal driver information such as calendar entries. We assume that this information is available by connecting the smartphone to the \ac{IVIS}. Below we present 5 ideas, on how TOR assistants can benefit from generative AI.

\paragraph{Interactive Scenarios}
Dynamic visual representations of scheduled \acp{TOR} can improve the usability of TOR assistants~\cite{Hollander.PreparingDrivers.2018}. Whereas current research focuses on relatively simple visualizations that are primarily focused on the timing or priority of the \ac{TOR}, we propose to use generative models such as \textit{DALL-E 3}\footnote{\url{https://openai.com/dall-e-3}} to generate dynamic scenarios that represent the upcoming driving situation. These scenarios can be displayed on the center stack screen as shown in \autoref{fig:teaser}\footnote{Some elements were generated using Adobe Illustrator's \textit{"Text to Vector Graphic"} feature:  \url{https://www.adobe.com/products/illustrator/text-to-vector-graphic.html}}, on the head-up display, or on the dashboard. For example, when approaching a highway exit, an image or video sequence of the exit can be displayed, prompting the driver to make a decision. While these scenarios can be used in combination with a direct prompt, they can also be used to subtly prime the driver for an upcoming TOR by displaying dynamic content on the screen in the periphery of the driver's focus.

\paragraph{Conversational Primers}
Research suggests that conversational voice assistants and priming techniques can help to build appropriate \ac{SA} and improve \ac{TOR} quality~\cite{Mahajan.ExploringBenefits.2021, Bai.UnlockingSafer.2023, SadeghianBorojeni.ReadingDriving.2018}. We argue that LLMs can further increase this potential as the system can engage the driver in natural but brief situation-pendent conversations about the upcoming route or driving scenario. For example, a question such as \textit{``Looks like we're getting off the highway in 10 minutes. Have you driven this route before?''} not only informs the driver of the upcoming TOR, but also indirectly prompts the driver to look at the road, thereby improving SA. This strategy can also be useful in situations where the system is uncertain whether a TOR will be triggered in the near future, as the driver may not even realize that the goal of the conversation was to redirect his attention to the road. This way, drivers won't be annoyed by false positives because they won't recognize them as such.

\paragraph{Context-Aware and Personalized TORs}
LLMs can provide concise, contextual descriptions or advice based on real-time sensor data. This information can be used, for example, to generate situation-based TORs: \textit{``We are approaching a construction zone on the right lane with a speed limit of 50 km/h, please take control''}. While current research suggests that context-aware warnings can lead to safer takeovers~\cite{Pakdamanian.EnjoyRide.2022}, these approaches can only detect predefined situations and are therefore limited to specific situations. By leveraging the vast amount of data generated by LLMs and object detection algorithms, TORs are no longer limited to these predefined degrees of freedom. Based on data from the cabin camera, TORs can be tailored not only to the driving situation, but also to the driver's state and current activity. The intelligent TOR assistant could tell the driver to put away the phone or tablet, arguing that there will be enough time after the construction zone to finish the current activity.

\paragraph{Subtle Nudges}
Nudging and persuasion can influence drivers to drive more economically~\cite{Meschtscherjakov.AcceptanceFuture.2009} and more safely ~\cite{Choudhary.NudgingDrivers.2022}. We argue that generative AI technology can be used to generate effective persuasion strategies for \acp{TOR}. Based on the driver's past behavior and responses, the generative AI can create tailored priming interventions or use the information gathered from past conversations to persuade the driver to be more aware or take over earlier. For example, the assistant might mention the driver's daughter's soccer game to subtly appeal to the driver's sense of responsibility not to get too distracted.

\paragraph{Ambient Scene Generation}
Ambient displays and audio cues are an effective measure to improve TOR quality~\cite{SadeghianBorojeni.AssistingDrivers.2016, SadeghianBorojeni.ReadingDriving.2018}. While current approaches are more or less explicit, we propose that based on the current or upcoming driving situations, an intelligent agent can generate situation-specific ambient scenes. For example, it could subtly change the tone of the infotainment system, or generate soft ambient sounds that resemble the road or traffic to subconsciously focus the driver's attention on the driving environment. The same applies for ambient lighting. The assistant could gradually synchronize the car's interior lighting with the outside environment and traffic scene. Dynamic lightning patterns based on passing cars or upcoming situations can be generated and visualized using ambient light technology. A slight change in brightness or hue can alert the driver's senses without the driver being aware of the change. 

\section{Proposed System Architecture}


\begin{figure}
  \includegraphics[width=\textwidth]{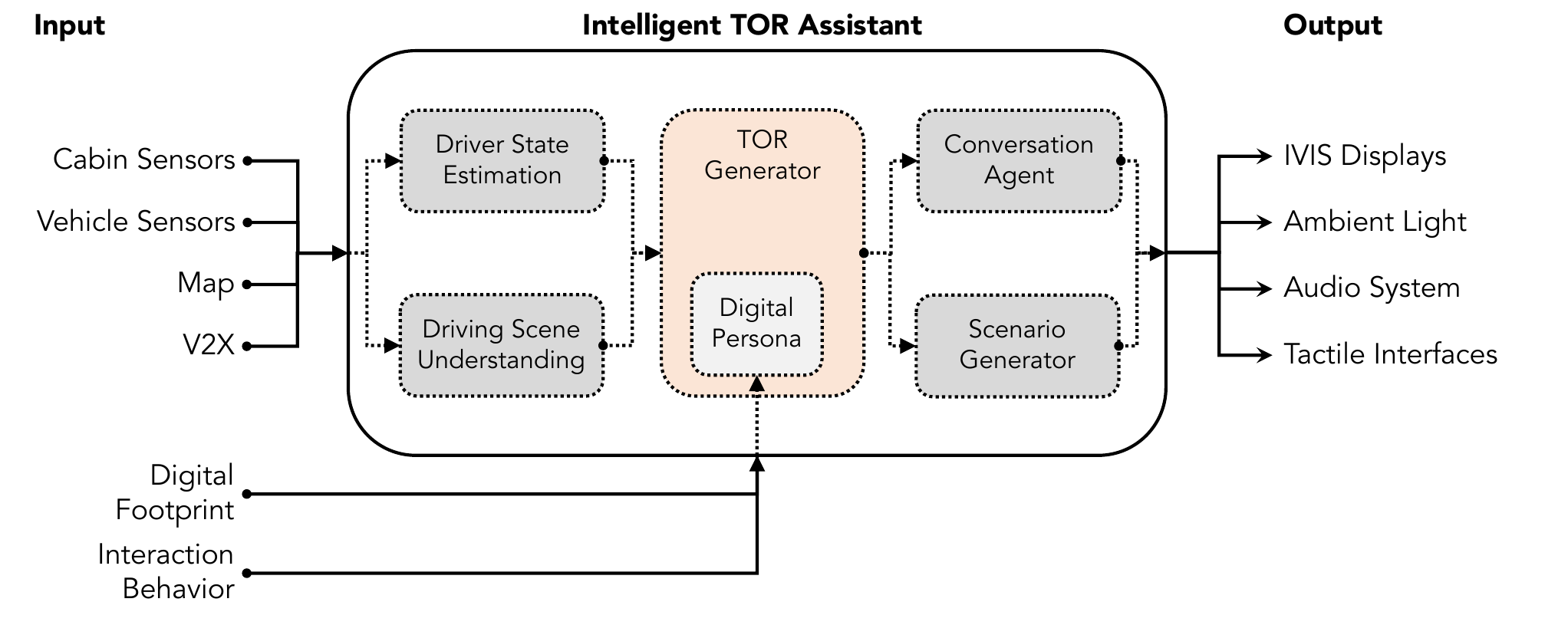}
  \caption{System Architecture}
  \label{fig:Architecture}
\end{figure}

\autoref{fig:Architecture} shows our proposed system architecture for an \textit{Intelligent TOR Assistant} that can apply the TOR strategies introduced above. To fully enable these strategies, an intelligent TOR assistant must create a holistic representation of the driving situation and the driver's state based on various types of inputs. 

We argue that in order to holistically assess the driver's state and understand the driving scene, the intelligent TOR assistant needs to access cabin sensors (e.g., cabin camera or cabin microphone), vehicle sensors (e.g., vehicle speed, steering wheel behavior, or automation status), map information (e.g., current location, future route, or traffic), and \ac{V2X} data (e.g., position and behavior of surrounding vehicles). This information is used to create a latent representation of the driver's state and the current driving scene, which is then used as input for the TOR generator.

Other inputs include the driver's digital footprint and interaction behavior. Digital footprint information describes all information available to the assistant about the driver's digital activities. This can include calendar entries or chat logs. Together with current and past interaction behavior (e.g., past conversations with the in-vehicle voice assistant or driving responses to TORs), this information forms the \textit{Digital Persona}. This digital persona is learned individually for each driver, enabling personalized predictions tailored to the driver's preferences and skills.

The \textit{TOR Generator} is the central unit of the intelligent TOR assistant. The TOR generator receives a representation of the current driver state and driving scene and combines this information with the digital persona to trigger context-sensitive, situation-aware, and personalized TORs. The TOR generator decides which of the above strategies is most appropriate for the current situation and triggers the \textit{Conversation Agent}, \textit{Scenario Generator}, or both. Based on the information received from the TOR generator, these two modules generate tangible outputs and communicate them to the driver via the appropriate output interfaces, the \ac{IVIS} displays, the ambient lighting, the audio system, and the tactile interfaces.

\section{Discussion and Conclusion}
We argue that key advantage of using generative AI for scheduled \acp{TOR} is subtlety and persuasion. The interactions should be smooth, non-intrusive, and feel natural so that the driver's SA is maintained without the driver actively realizing that they're being assisted. The goal is not to make the driver dependent on the Intelligent TOR Assistant, but to use the new opportunities that generative AI methods provide to enhance the collaboration between driver and the automated driving system. While subtle cues can help drivers to maintain an appropriate level of SA, LLMs can also be used to generate eloquent and meaningful prompts that persuade the driver to be more attentive. Incorporating personal and situational information could not only improve in-situ TOR quality, but also change driver behavior in the long run. 

For all of the strategies presented in this position paper, it is important to emphasize that TORs are safety-critical. Choosing an inappropriate modality or providing false or inaccurate information can have fatal consequences. This needs to be considered in future work, especially in light of current vulnerabilities of generative models such as hallucination, bias, and lack of explainability. In addition, the question of how to ensure that approaches using generative AI methods comply with regulations needs to be answered. Due to their non-deterministic nature, they can't be evaluated against standardized datasets to assess whether they are \textit{``good enough''} to be used for safety critical applications\footnote{Not to say that the question of what is \textit{``good enough''} when it comes to automated driving has been answered yet.}.

While some of the above strategies may seem dystopian at the time of this writing, a digital assistant that is intimately aware of user preferences and behaviors and can carry on a conversation as naturally as a human counterpart may be technically possible and socially acceptable in just a few years. However, research suggests that conversational agents that seem too human don't necessarily drive adoption. In fact, they may deter people from using the technology~\cite{Fernandes.UnderstandingConsumers.2021}. Thus, implementing strategies such as the \textit{Subtle Nudges} strategy is a challenging endeavor and more research is needed to enable systems such as the one presented in this position paper.

\bibliography{AUIs}

\end{document}